\documentclass[traditabstract]{aa}
\usepackage{lineno}
\usepackage[varg]{txfonts}
\usepackage{graphicx}
\usepackage{xcolor}
\usepackage{natbib}
\usepackage{amsmath,amssymb}
\usepackage{hyperref}
\usepackage{rotating, bm}

\begin{document} 

\title{
The uncertainty of magnetic fields in 3D non-local thermodynamic equilibrium  inversions
}

\titlerunning{Uncertainties in 3D inversions}

\author{
Ji\v{r}\'{\i} \v{S}t\v{e}p\'an\inst{1}
\and
Tanausú del Pino Alemán\inst{2,3}
\and
Andrés Vicente Arévalo \inst{2,3,4}
}
\authorrunning{\v{S}t\v{e}p\'an et al.}

\institute{
Astronomical Institute of the Academy of Sciences, Ond\v{r}ejov, Czech Republic. \email{jiri.stepan@asu.cas.cz}
\and
Instituto de Astrof\'{\i}sica de Canarias, E-38205 La Laguna, Tenerife, Spain
\and
Departamento de Astrof\'{\i}sica, Universidad de La Laguna, E-38206 La Laguna, Tenerife, Spain
\and
Leibniz Institut für Sonnenphysik (KIS), Georges-Köhler-Allee 401a
79110 Freiburg, Germany.
}

\date{Received XXXX; accepted XXXX}

\abstract{We describe our approach to solving the problem of ensuring the solenoidality of the magnetic field vector in 3D inversions, as well as the estimation of the uncertainty in the inferred magnetic field. The solenoidality of the magnetic field vector is often disregarded in the inversion of spectropolarimetric data due to limitations in the traditional 1D inversion techniques. We propose a method for ensuring the solenoidal condition in 3D inversions based on our mesh-free approach. The increase in dimensionality with respect to the 1D inversion techniques is such that some of the traditional methods for determining the uncertainties become unfeasible. We propose a method based on a Monte Carlo approach for determining the uncertainty of the magnetic field inference. Due to the physics of the problem, we can compute the uncertainty while only increasing the total required computational time by a factor of about 2. We also propose a metric for quantifying the uncertainty that thus describes the degree of confidence of the magnetic field inference. Finally, we perform a numerical experiment to demonstrate the feasibility of both the method and the metric proposed to quantify the uncertainty.
}


\maketitle

\section{Introduction
\label{sec:intro}}

The magnetic field plays a key role in the solar upper atmosphere, driving its dynamics and activity. Spectropolarimetry and inversion techniques are powerful tools for unraveling the properties of these magnetic fields \citep[see, e.g., the review by][]{2017SSRv..210..109D}. However, inferring the 3D distribution of the magnetic field vector from spectropolarimetric observations requires sophisticated methods that account for the intricate interplay of the physical effects that produce the polarization signals, and thus it remains a significant challenge.
Spectral lines in the upper solar atmosphere typically form out of the local thermodynamic equilibrium (NLTE) conditions,
and the information carried in their spectropolarimetric profiles, influenced by the Zeeman and Hanle effects, is much richer than that of the photospheric lines forming in, or close to, local thermodynamic equilibrium (LTE) conditions.
On the one hand, the nonlocal and nonlinear coupling introduces considerable challenges for both the physical interpretation and numerical modeling of these spectropolarimetric signals. On the other hand, it offers a framework for inferring the magnetic field vector and other quantities throughout the entire 3D geometry in a physically consistent manner.

It has long been recognized that traditional pixel-by-pixel inversion techniques, while providing significant insight into the magnetic field, are insufficient for the accurate and robust interpretation of spectropolarimetric observations because they fail to account for the spatial coupling of physical quantities across different plasma regions \citep[see, e.g.,][]{2012A&A...548A...5V}. Several recent studies have incorporated spatial coupling into inversion techniques to address this limitation. \cite{2015A&A...577A.140A} introduced sparsity regularization and showed its potential by implementing it in a Milne-Eddington inversion. \cite{2019A&A...631A.153D} considered the spatial coupling due to the combination of observations with different spatial sampling and instrumental point spread functions and presented their method in a Milne-Eddington inversion. The potential of machine learning techniques such as neural fields has also been explored \citep{2025A&A...693A.170D}, and such techniques have been implemented in the spatially coupled application of the weak-field approximation of the Zeeman effect.

Although these approaches entail a significant advance toward the more robust inference of the 3D spatial distribution of the magnetic field vector, they remain limited.\ They cannot easily be applied to upper-chromosphere spectral lines because they only consider the spatial coupling of thermodynamic quantities, neglecting radiative interactions between matter in different regions in the 3D plasma. Forward-modeling studies, such as those using the PORTA code \citep{2013PORTA}, have demonstrated the importance of including radiative coupling to capture the full complexity of the radiative transfer (RT).  Without accounting for these effects, traditional methods may not be able to fully resolve the intricacies of 3D magnetic field topologies, remaining mostly restricted to the estimation of the line-of-sight (LOS) component of the magnetic field in strongly magnetized active regions.

To consider the spatial coupling in full, \citet{2022A&A...659A.137S} proposed a new framework for the 3D inversion of spectropolarimetric data, allowing for the inclusion of all relevant physical processes, including RT. The method is formulated as a mesh-free and unconstrained minimization problem that includes stochastic ingredients. The unconstrained nature of the problem avoids the need for self-consistency (i.e., the solution of the forward problem at every step of the inversion process), which instead appears in the loss function as a regularization. The method's stochastic nature helps avoid local minima and saddle points in the global loss function. Additional constraints, such as preservation of the magnetic field solenoidality, $\nabla\cdot\vec B=0$, and hydrostatic equilibrium, can be imposed via regularization terms in the loss function. We have recently implemented these ideas in the POLARIS code, which we used throughout this investigation.

Another critical aspect of the inference of the magnetic field is that of the uncertainties. Not only is the inversion problem ill-posed, but observations have limited spectral, spatial, and temporal resolution and are affected by noise. Without a proper quantification of the uncertainty, the inference is just one of the many possible solutions compatible with the data. A point estimated error can be derived from the response functions\citep[][]{1997ApJ...491..993S}, which provide an indication of how well constrained a node value is relative to the others \citep[][]{2018A&A...617A..24M}. However, while straightforward and relatively computationally cheap in a typical 1D inversion, such an approach is impractical in our mesh-free 3D method. A much better estimation of the uncertainty can be achieved using statistical techniques, such as Bayesian methods \citep[][]{2007A&A...476..959A}. While it has been shown that neural networks can help with the acceleration of these time-consuming techniques \citep{2022A&A...659A.165D}, such an approach is impractical in 3D. A different statistical approach consists of performing a significant number of inversions and introducing some randomization in the initial model and/or in the spectropolarimetric profiles \citep[][]{WestendorpPlaza2001ApJ,2016LRSP...13....4D,SainzDalda2023ApJ}, with the uncertainty given by the standard deviation of the whole ensemble of inversions. While a priori this approach also seems impractical for the very time-consuming inversions, we show how it can be implemented to determine the uncertainty of the magnetic field inference under certain reasonable assumptions.

In Sect.~\ref{sec:solen} we introduce a method capable of ensuring the exact preservation of the solenoidal nature of the magnetic field throughout the entire domain, which allows us to avoid the addition of the solenoidality regularization term in the loss function. This is made possible by the way that our 3D inversion code, POLARIS, treats the model variables. In Sect.~\ref{sec:err} we address the challenging task of estimating the uncertainties of the inferred magnetic field vector. This is particularly challenging in 3D models due to the large number of model parameters. We present a simple method for estimating the uncertainties of the magnetic field vector, demonstrating how to derive its probability distribution at any point in the model. Additionally, we propose a quantitative approach for assessing the relative uncertainty of the inversion of the magnetic field across different regions, providing a practical framework for the visualization of the magnetic field in 3D geometry. In Sect.~\ref{sec:experiment} we apply the abovementioned methods in a numerical experiment in which we invert a known magnetic field in a 3D volume. Finally, our conclusions are
summarized in Sect.~\ref{sec:concl}.

\section{Exact preservation of the magnetic field solenoidality
\label{sec:solen}}

\begin{figure}
\centering
\includegraphics[width=0.8\columnwidth]{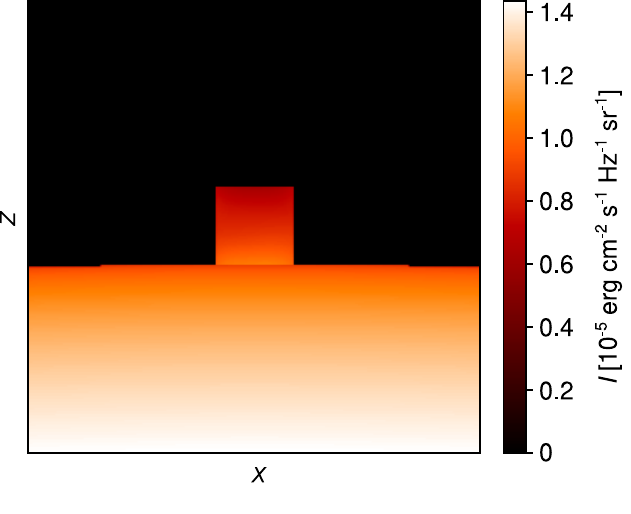}
\caption{Line-center image of the $5\times 5\times 5\;\mathrm{Mm}^3$ model prominence just above the solar limb used in this paper, observed from Earth with a FOV of $20''\times 20''$, at the line center of the academic line at 5000~\AA. The disk illumination is given by the disk center intensity and center-to-limb variation in \cite{1976asqu.book.....A}.}
\label{fig:fov-big}
\end{figure}

\begin{figure}
\centering
\includegraphics[width=0.8\columnwidth]{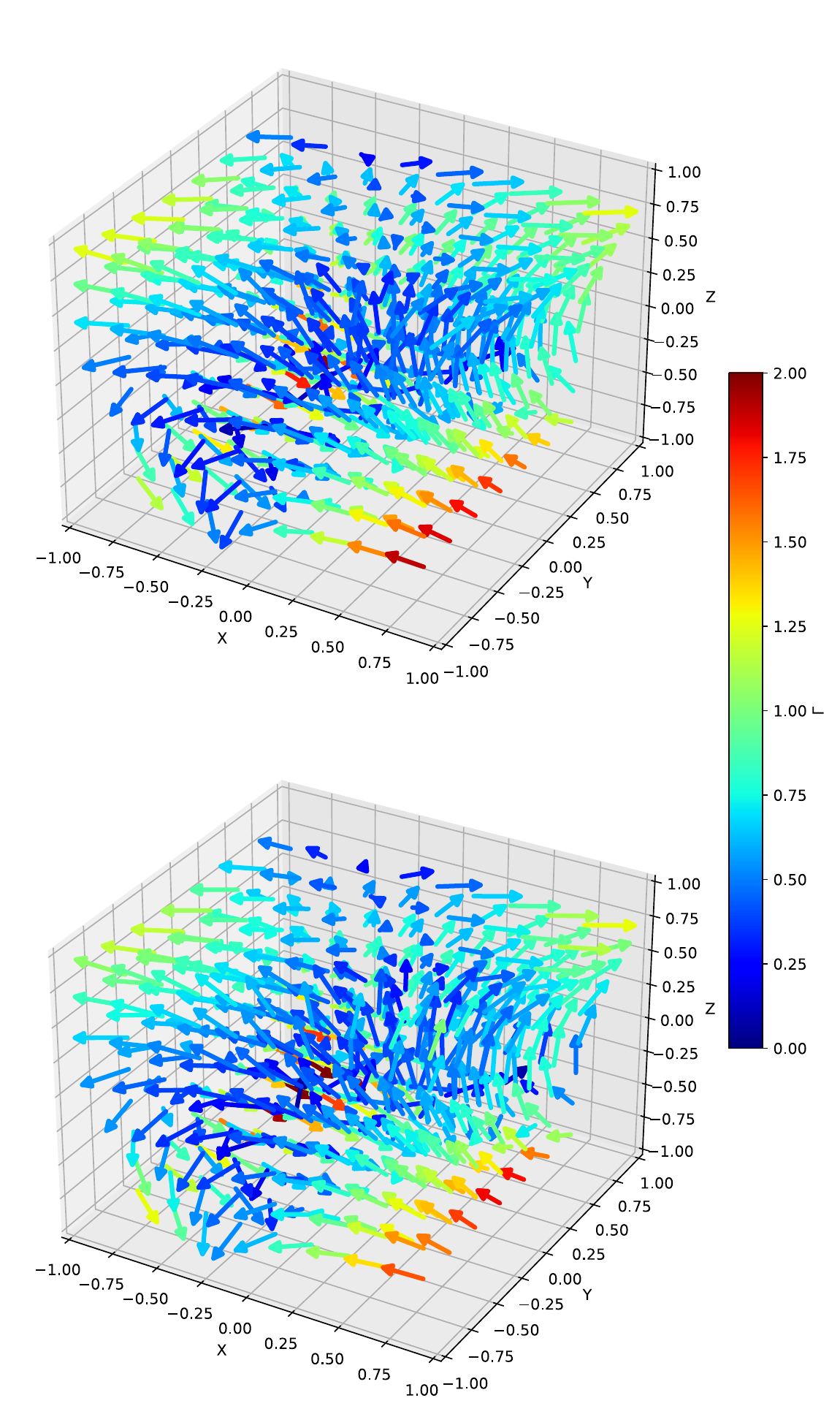}
\caption{Magnetic field vectors in the true (top panel) and inverted (bottom panel) models. The magnetic field strength is normalized to the critical Hanle field of the spectral line used in Sect.~\ref{sec:fwd}, $\Gamma=B/B_{\rm H}$ with $B_{\rm H}=284$~G.
}
\label{fig:3dfield}
\end{figure}

\begin{figure}
\centering
\includegraphics[width=0.8\columnwidth]{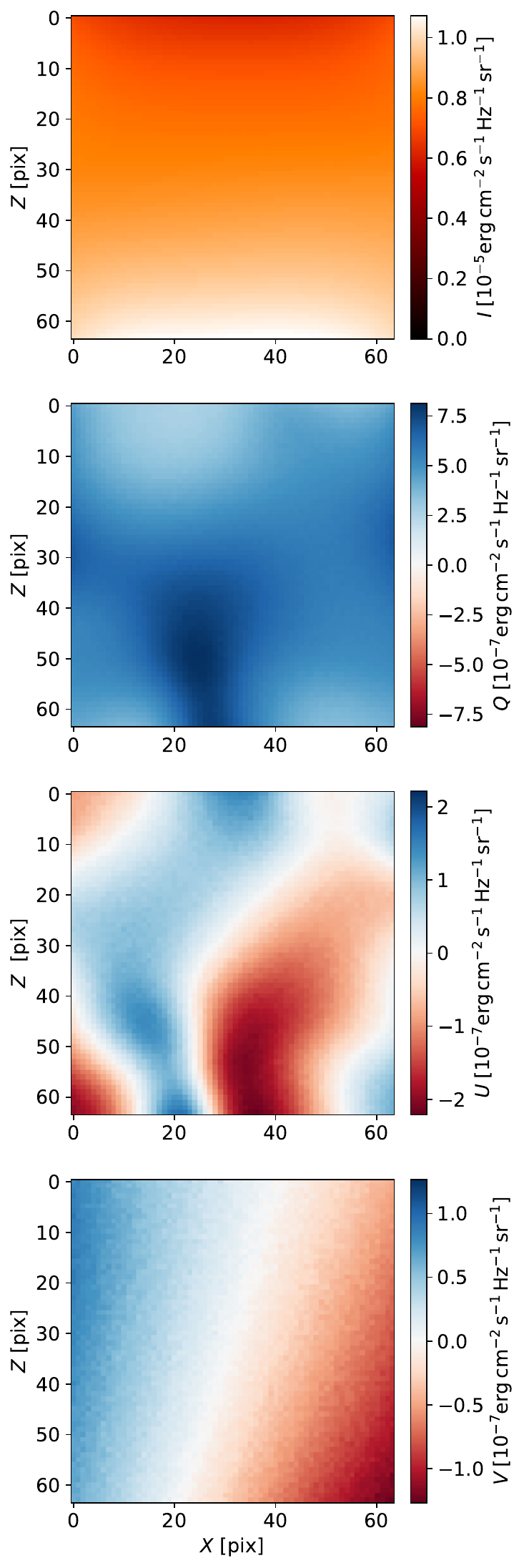}
\caption{
Synthetic Stokes parameters resulting from the forward modeling in the prominence model shown in
Fig.~\ref{fig:fov-big}, with the $64\times 64$ pixels FOV including only the prominence, and with
added Gaussian noise with $\sigma = 2\times10^{-9}$~erg\,cm$^{-2}$\,s$^{-1}$\,Hz$^{-1}$\,sr$^{-1}$.
The top, second, and third panels show the intensity and Stokes $Q$ and $U$ for the line center
at 5000~\AA, respectively, while the bottom panel shows Stokes $V$ at the wavelength of the maximum signal at $-0.14$\,\AA\ from the line center.
}
\label{fig:fov}
\end{figure}

\begin{figure}
\centering
\includegraphics[width=0.7\hsize]{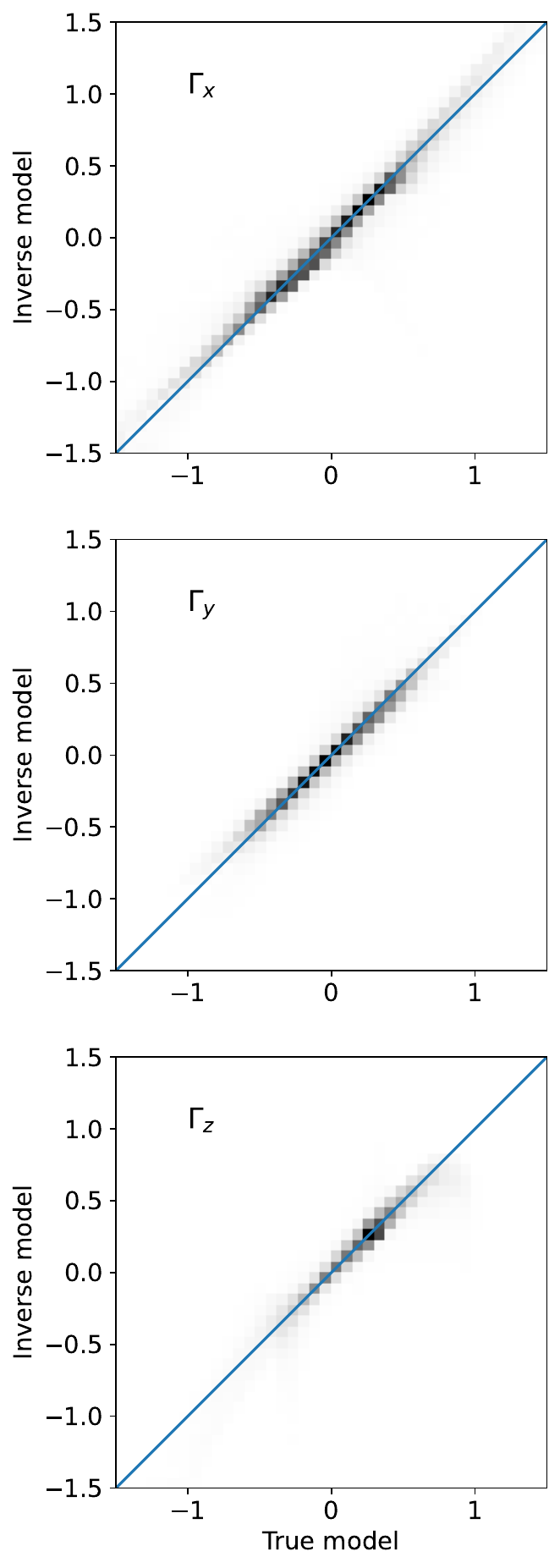}
\caption{Inferred magnetic field components versus their true value in the original model. The
different panels correspond to the magnetic field vector component along the $x$, $y$, and $z$
axes in the model, from top to bottom, respectively. The figure considers $10^5$ samples in
random positions in the 3D domain. All magnetic field components are normalized to the critical
Hanle field, $\Gamma_i=B_i/B_{\rm H}$.
}
\label{fig:corr}
\end{figure}

\begin{figure}
\centering
\includegraphics[width=0.7\hsize]{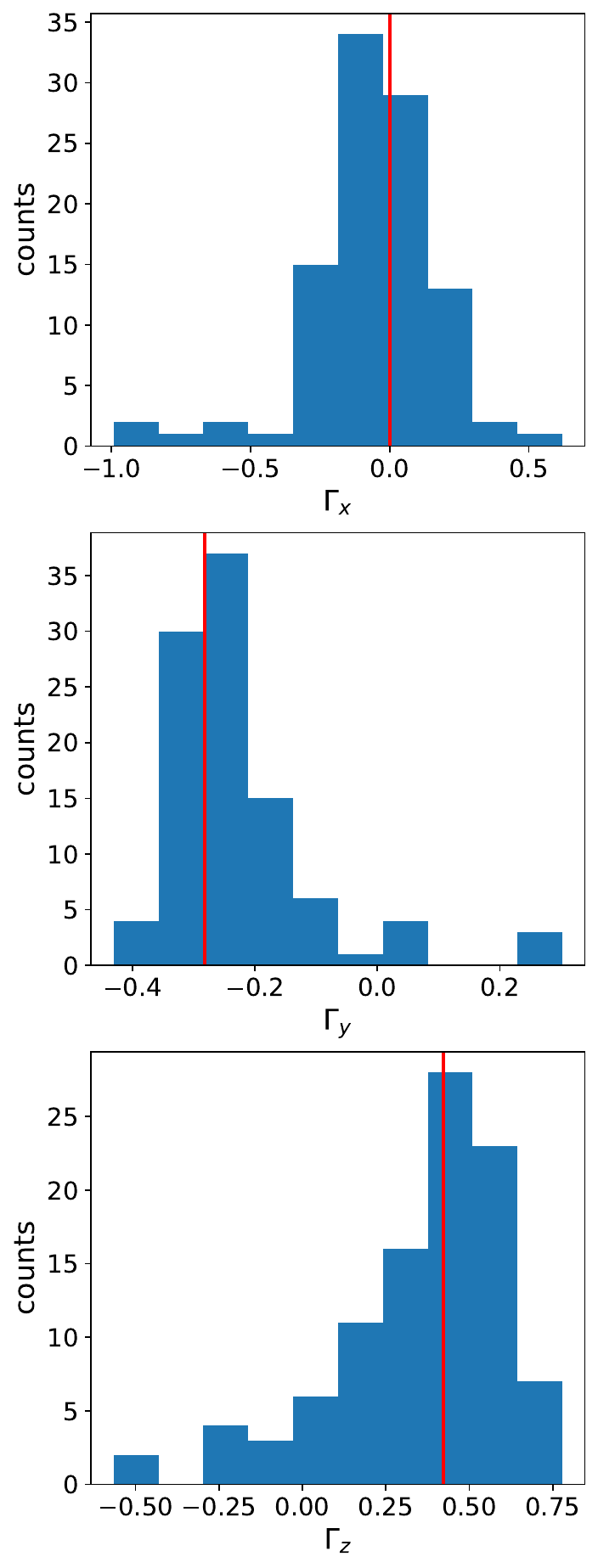}
\caption{Histogram of the magnetic field components at the position with normalized coordinates
$\vec r = (0, 0, 0.5)$ in the 3D domain for 100 different inverted models. The
different panels correspond to the magnetic field vector component along the $x$, $y$, and $z$
axes in the model, from top to bottom, respectively. The vertical red line indicates the
true value of the magnetic field vector. The relative uncertainty for each component at this
position, as defined in Eqs.~\eqref{eq:udef} and \eqref{eq:uxyzdef}, are $U_x=0.97$, $U_y=0.50$,
$U_z=0.60$, and $U_{xyz}=0.66$.
}
\label{fig:hist}
\end{figure}

\begin{figure*}
\centering
\includegraphics[width=1\hsize]{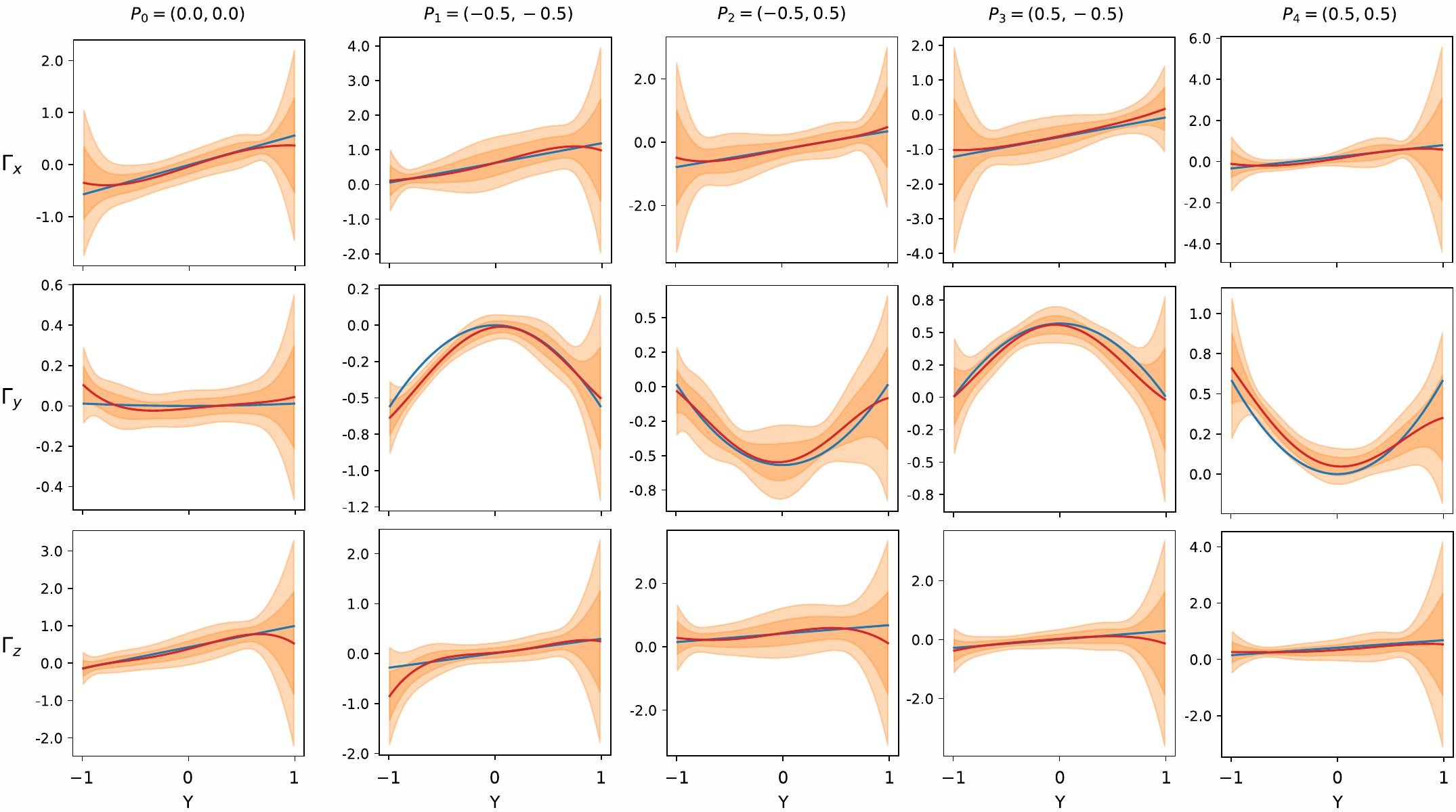}
\caption{
Variation in the magnetic field vector components along the LOS. From top to bottom, we show
the magnetic field vector component along the $x$, $y$, and $z$ axes in the model, respectively.
The different columns correspond to different positions in the FOV, whose location in normalized
coordinates, $P_i=(X,Z)$, is indicated to the left of each row. The blue curve shows the true 
magnetic field and the red curve the result of the inversion. The dark and light orange-shaded
regions indicate the 68\,\% and 95\,\% confidence intervals of the inversion, respectively.
}
\label{fig:los}
\end{figure*}

\begin{figure*}
\centering
\includegraphics[width=\hsize]{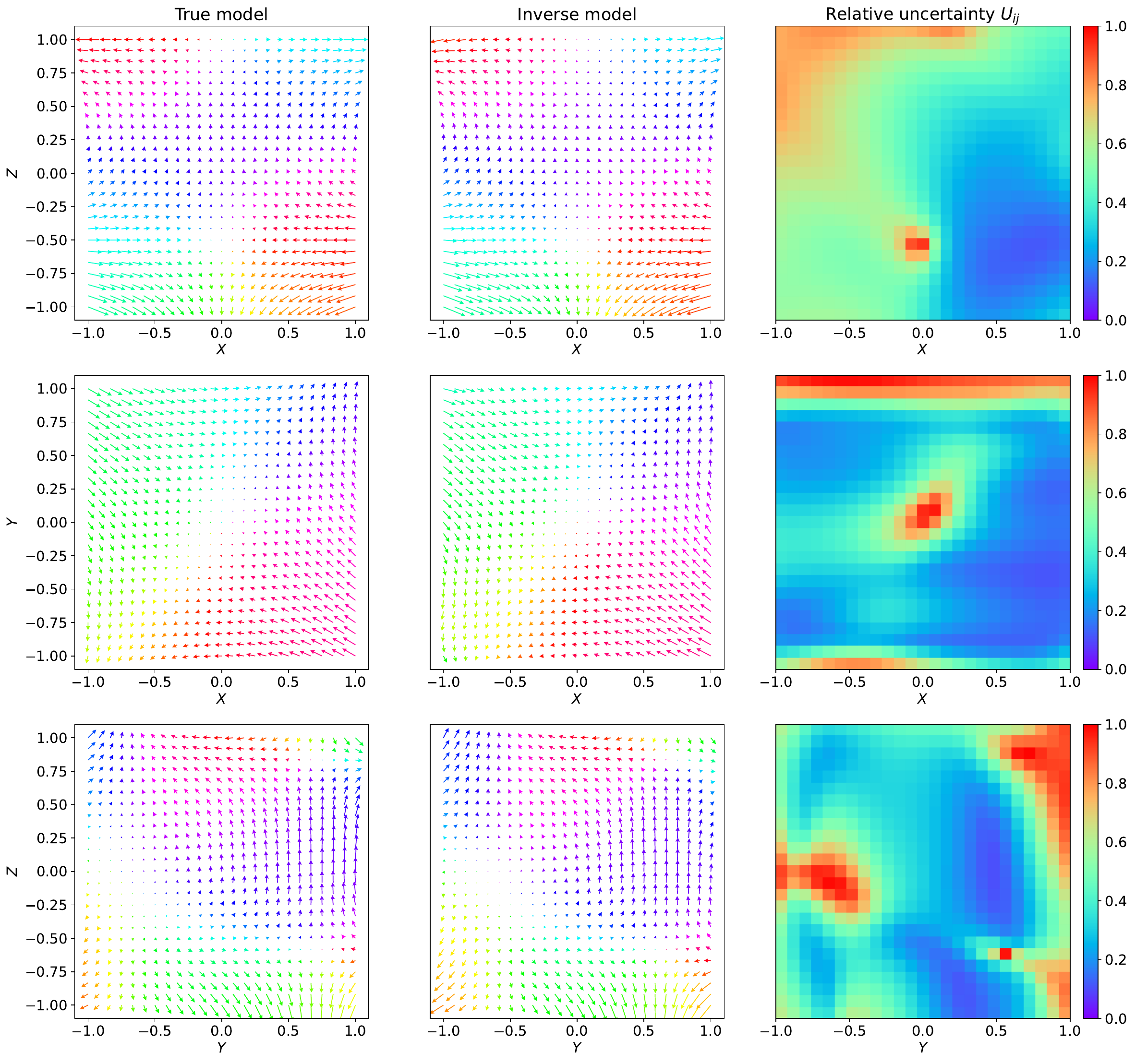}
\caption{Projection of the magnetic field vector in the $X$--$Z$ (top row), $X$--$Y$ (middle row), and
$Y$--$Z$ (bottom row) planes intersecting the center of the 3D domain. The first and second
columns show the projection of the true and the inverted magnetic field vectors, respectively.
The right column shows the uncertainty of the inverted magnetic field as defined in
Eq.~\eqref{eq:uxyzdef}, but accounting only for the two relevant components in each projection.
The color of the vector arrows indicates their orientation on the projected plane.
}
\label{fig:slices}
\end{figure*}

The Maxwell equation $\nabla \cdot \vec{B} = 0$ represents a fundamental condition that must be satisfied in any physical model. This condition, known as solenoidality, is naturally imposed in 3D magnetohydrodynamic (MHD) simulations. However, it is typically neglected in NLTE inversion theory because standard 1D inversions do not readily accommodate it. Nevertheless, certain techniques designed to minimize the current density and $\nabla \cdot \vec{B}$ in the photosphere have been employed in the LTE regime to solve the $180^\circ$ ambiguity \citep[e.g.,][]
{2006SoPh..237..267M}. Recently, \cite{2024A&A...687A.155B} introduced a method for incorporating this constraint by computing one of the magnetic field vector components from the other two to satisfy $\nabla \cdot \vec{B} = 0$.

In \citet{2022A&A...659A.137S}, we proposed a comprehensive 3D NLTE inversion strategy that
seamlessly incorporates the solenoidality condition as a regularization term in the global loss
function. This approach can also include other physical constraints, such as the equation of
continuity or hydrostatic equilibrium. While certain constraints, like hydrostatic equilibrium,
are approximate due to the dynamic nature of solar plasma, the solenoidality condition is exact
and should be strictly enforced, irrespective of the physical model employed.

In the code POLARIS, we implement the ideas presented in \citet{2022A&A...659A.137S} using a
mesh-free approach, eliminating the need for a traditional grid. This design facilitates the
representation of a quantity $\theta_m$ at a spatial point $\vec{r} \equiv (x, y, z)$ through
the expression
\begin{equation}
\theta_m(\vec{r}) = \mu(m) \sum_{ijk} c_{ijk}(m) \phi_i(x) \phi_j(y) \phi_k(z)\,.
\label{eq:qdef}
\end{equation}
Here, the normalization factor $\mu(m)$ ensures the correct physical dimension and normalization of
$\theta_m$, and the $\phi_n(\cdot)$ functions --- specifically, Chebyshev polynomials of the first order,
$T_n(x)$, in the current version of POLARIS --- form an orthogonal functional basis in 3D space. This
representation allows for solutions whose smoothness is controlled by the maximum order of the basis
functions, effectively regularizing inverse solutions. The coefficients $c_{ijk}(m)$ are the model
parameters to be inferred during the inversion process. A key advantage of this implementation is
that quantities are represented as smooth and differentiable functions throughout the whole domain.

The magnetic field vector $\vec{B}$ can always be expressed as the curl of a vector potential,
$\vec{B} = \nabla \times \vec{A}$. This formulation in terms of a vector potential inherently
ensures solenoidality because the divergence of a curl is zero for any vector field $\vec{A}$.
In the approach presented in this work, we invert the vector potential $\vec{A}$ instead of the
magnetic field vector $\vec{B}$. This ensures the exact preservation of solenoidality. This method
has been previously employed in MHD simulations, such as those by \citet{manabu09}. The vector
potential $\vec{A}$ is not uniquely defined due to its gauge invariance, allowing the addition
of any curl-free field without affecting $\vec{B}$. We do not impose any gauge-fixing conditions,
such as the Lorenz or Coulomb gauges, which could reduce the number of model parameters, and instead
we retain all the degrees of freedom of $\vec A$ in the model. This makes the implementation easier
and provides the stochastic iterative algorithm with more pathways to explore the parameter space
to achieve the global minimum of the loss function \citep[see][for details on the stochastic
iterative process]{2022A&A...659A.137S}. Since all quantities in the model are represented by
smooth and differentiable functions across the whole domain, the calculation of partial derivatives
required for the curl operator is straightforward.

This approach, while maintaining the same computational demands as the traditional regularization method, leads to configurations in the solution space that inherently satisfy the solenoidality condition. This has two key benefits: first, it guarantees the physical consistency of the model regarding $\nabla \cdot \vec{B} = 0$, which is strictly satisfied. Second, it eliminates the need to introduce a weighting factor for an additional penalty term in the optimization process. However, a minor trade-off is that achieving a given order of smoothness for $\vec{B}$ requires the basis functions for $\vec{A}$ to be one order higher due to the curl operator.

\section{Estimation of the error of the inverted magnetic field
\label{sec:err}}

The estimation of uncertainties in the determination of the magnetic field vector is critical in any inverse problem, as its solution is never unique. Spectral lines, while providing essential diagnostic information, are inherently limited by noise in their profiles, as well as the spectral and spatial resolution of the observations. Without quantifying the associated uncertainties, the solutions derived from these inversions only represent one of many possible configurations, which undermines their scientific reliability. Developing robust error estimation techniques is therefore essential for data interpretation. As mentioned above, while several methods for quantifying parameter uncertainties have been proposed in 1D scenarios, their extension to 3D inversions poses significant challenges related to the increased complexity of the parameter space, computational cost, and the way the physical quantities are represented. Nonetheless, some approaches are sufficiently general to remain applicable even within the complexity of 3D geometries.

Exploring the entire parameter space to identify consistent models that fit the observations would, in principle, allow for the construction of probability distributions for individual parameters and the derived physical quantities. However, in the context of 3D inversions, the vast number of model parameters makes such an exhaustive exploration computationally impossible. We address this challenge by narrowing our investigation to the magnetic field, which offers a more tractable problem compared to other variables. This simplification is justified because changes in the magnetic field significantly affect the goodness-of-fit parameter $\chi^2$, while typically having a negligible impact on the thermal structure of the inverted model. Specifically, fixing the thermal structure and the $J^K_Q$ tensors, i.e., our atomic-like variables as defined in the \citet{2022A&A...659A.137S} framework, results in minimal changes to the NLTE consistency of the model under variations in the magnetic field. While this does not apply to the $J^1_Q$ components, which depend exclusively on the circular polarization, their impact on the problem is generally minor. Even in cases in which these tensors impact the solution, the proposed approach remains effective for rough error estimation. By taking advantage of this property, we employed the well-established Monte Carlo (MC) strategy outlined in Sect.~15.6.2 of \citet{press2007numerical}. This approach enables a systematic and computationally efficient estimation of uncertainties specific to the magnetic field. The algorithm can be summarized as:
\begin{enumerate}
\item Start from the converged inverse model.
\item Synthesize the field of view (FOV) spectra from this model, creating synthetic data that will serve as the reference for the uncertainty estimation process.
\item Repeat the following steps $N$-times:
\begin{enumerate}
    \item Add random noise with the same distribution as in the observation to the synthetic FOV spectra.
    \item Randomize the vector potential $\vec A$ by randomizing its coefficients $c_{ijk}$, sampling them from the normal distribution $\mathcal{N}(0,\sigma^2)$ with a sufficiently large variance $\sigma^2$.
    \item Solve the inverse problem to recover the vector potential $\vec A$ while keeping the rest of the model parameters fixed, that is, the thermal structure and the $J^K_Q$ tensors.
    \item If the inversion problem converges, keep the resulting model for the vector potential.
\end{enumerate}
\end{enumerate}
As a result, we obtain up to $N$ of the $\vec A$ vector potential configurations that are consistent with the observations.\footnote{Any inverted model from an inversion failing to converge should be excluded from the analysis to ensure the reliability of the results.} Each configuration corresponds to a magnetic field $\vec B=\nabla\times\vec A$, whose Cartesian components, $B_i(\vec r)$, can be calculated at any point $\vec r$ within the domain. The resulting ensemble of magnetic fields provides a distribution that can be used to estimate the uncertainties as discussed in the following section.
We note that sampling the $c_{ijk}$ coefficients from a normal distribution ensures that the distribution of initial $A_i$ components at every point is also sampled from a normal distribution with zero mean.

The proposed algorithm is efficient for several reasons. First, it does not require checking for NLTE consistency, avoiding the use of pilot points, their associated long characteristics, and the need to calculate RT along them. Second, based on our experience, a smaller subset of random pixels can be used for estimating $\chi^2$ when compared to the full NLTE inversion without compromising the convergence rate. Lastly, since the only free parameters are those associated with the vector potential $\vec A$, the total number of parameters is significantly reduced, typically by about a factor of 5 in relatively simple models like the example in Sect.~\ref{sec:experiment}, and by a factor of 10 or more in more complex models. As a result, the CPU time required to solve the inverse problems in this algorithm can be roughly estimated to be about 1\,\% or less of the time needed for a full NLTE inversion. This estimate shows that about 100 inverse models can be obtained in about the same CPU time as the full NLTE inversion.\footnote{This number needs to be taken as an order-of-magnitude estimate because it depends on a number of factors, such as the number of pilot points in the full inversion, the complexity of the atomic models, etc.} Such a number of samples can already be used for the statistical analysis of the uncertainties.

\subsection{Quantification of uncertainties
\label{ssec:qunc}}

The most comprehensive representation of the magnetic field uncertainty at a specific spatial point in the 3D domain is the distribution of the magnetic field vector obtained using the MC method described above. If this distribution is unimodal and well approximated by a 3D normal distribution, the uncertainty can be effectively characterized by the standard deviation, $\sigma(B_i(\vec r))$, of the marginal distributions for each Cartesian component of the field.

Visualizing the inferred magnetic field is clearly more challenging in 3D than in 1D inversions, particularly when incorporating error bars. It is thus useful to define a scalar quantity to quantify the relative reliability of the inversion across different regions in the domain. One such measure is the coefficient of variation, $C_i=\sigma(B_i(\vec r))/\mu(B_i(\vec r))$, where $\mu(B_i(\vec r))$ stands for the mean value. The absolute value of this quantity decreases as the certainty of the inversion increases. However, a limitation of this measure is that it can take both positive and negative values and that it can become arbitrarily large. To overcome this issue, we defined the relative uncertainty using the standard deviation and the root mean square as
\begin{equation}
U_i(\vec r)=\frac{\sigma_i(\vec r)}{\sqrt{\langle B_i^2(\vec r)\rangle}} .
\label{eq:udef}
\end{equation}
It is straightforward to demonstrate that $U_i$ ranges from 0 (indicating complete certainty) to 1 (indicating complete uncertainty). Hereafter, we drop the dependence of $U$, $\sigma$, and $B_i$ on the position, $\vec r$, for notational convenience. A single metric for the entire magnetic field vector can then be defined as
\begin{equation}
U_{xyz}=\sqrt{\frac{\sigma_x^2+\sigma_y^2+\sigma_z^2}{\langle B_x^2\rangle+\langle B_y^2\rangle+\langle B_z^2\rangle}} .
\label{eq:uxyzdef}
\end{equation}
In Sect.~\ref{sec:experiment} we show an application of the proposed algorithm and of these definitions in a numerical example.

It is helpful to establish a threshold value of $U_i$ that indicates acceptable reliability. Assuming that the $B_i$ component follows a normal distribution with mean $\mu_i$ and standard deviation $\sigma_i$, $U_i$ can be expressed as
\begin{equation}
U_i=\left(1+\frac{\mu_i^2}{\sigma_i^2}\right)^{-\frac 12}\,.
\end{equation}
The ratio $|\mu_i/\sigma_i|$ quantifies the number of standard deviations by which the distribution is shifted from zero. A threshold of $|\mu_i/\sigma_i| = 1$, corresponding to $U_i \approx 0.7$, can be considered an acceptable level of uncertainty, depending on the application. For the relative uncertainty of the entire vector, it follows that
\begin{equation}
U_{xyz}=\left(1+
\frac{\mu_x^2+\mu_y^2+\mu_z^2}{\sigma_x^2+\sigma_y^2+\sigma_z^2}
\right)^{-\frac 12}\,.
\end{equation}

\subsection{Mean magnetic field model
\label{ssec:fwdmodel}}

According to Eq.~(\ref{eq:qdef}), we can write any Cartesian component of the vector potential at a given point $\vec r$ as
\begin{equation}
A_i(\vec r) = \mu(A_i) \sum_{klm} c_{klm}(A_i)\phi_k(x)\phi_l(y)\phi_m(z)\;.
\end{equation}
We can then calculate the mean of this quantity from the ensemble of inverse models,
\begin{equation}
\langle A_i(\vec r)\rangle = \mu(A_i) \sum_{klm} \langle c_{klm}(A_i)\rangle \phi_k(x)\phi_l(y)\phi_m(z)\,,
\end{equation}
where the averaging $\langle c_{klm}(A_i)\rangle$ is performed on the set of the converged inverse
problems. Thanks to the linearity of the problem, we can calculate the mean solution for the magnetic
field vector as
\begin{equation}
\langle \vec B(\vec r)\rangle = \nabla\times \langle \vec A(\vec r)\rangle\,,
\end{equation}
which still preserves the solenoidal condition. We note that other fields constructed from the statistical ensemble may not preserve solenoidality. In particular, the median of the Cartesian components of both $\vec A$ and $\vec B$ fields is generally not an analytical function.

The procedure described above is meaningful to characterize the uncertainty only when the distribution of the magnetic field vector $\vec B$ is unimodal, i.e., when the inversion does not yield multiple ambiguous solutions. In some instances, the mean model can offer an improvement over the original inverse model. This is because while individual inversions may significantly overestimate or underestimate the magnetic field at certain spatial points, the mean field can help reduce the impact of outliers.

\subsection{Note on ambiguities in 3D problems
\label{ssec:amb}}

Ambiguities in the solar spectropolarimetric inverse problem arise from the inherent degeneracies in the mapping of the magnetic field vector to the observed polarization signals. These ambiguities can appear for both the Zeeman and Hanle effects \citep[see, e.g.,][]{BLandiLandolfi2004,Asensioetal2008}. Both effects depend on the geometry of the magnetic field and the LOS, and in the case of the Hanle effect also on the illumination conditions. Different magnetic field geometries in specific configurations produce indistinguishable polarization signatures in the emergent radiation, leading to multiple plausible solutions for the same observed data.

Among the primary causes of these ambiguities are symmetries in the RT problem under simplified assumptions. For example, in the constant-property slab approximations, the RT equation is solved assuming that local plasma conditions and the cylindrically symmetric illumination dominate the formation of the spectral line \citep{Asensioetal2008}.
These assumptions neglect the radiative coupling between
distinct regions of the slab, which can
break the symmetry of the problem. Furthermore, as for relatively optically thin lines, it is often assumed that
the sensitivity of the emergent polarization is limited to a single or a few particular magnetic field vectors (in the case of the so-called multicomponent models).

However, when considering an optically thick medium with full 3D NLTE RT, we expect the symmetries of the problem to be significantly reduced. In this regime, the radiation field is strongly coupled to both the local and nonlocal thermodynamic and magnetic conditions. NLTE effects account for departures from equilibrium in the population and quantum coherence of atomic levels, which are influenced by the anisotropy and geometry of the radiation field. This coupling introduces a directional dependence to the emergent radiation sensitive to variations in the magnetic field and its surrounding environment. The inclusion of 3D RT further enhances this sensitivity by capturing the spatial variations in the illumination, breaking the symmetry found in simpler 1D models. In conclusion, 3D NLTE RT in optically thick media has the potential to resolve or mitigate many of these degeneracies. This approach thus generally provides a more unique mapping of the magnetic field to the observed polarization signals. While it may not completely eliminate all ambiguities, especially in cases with significant noise levels, it offers a promising path toward reducing their prevalence and improving the reliability of solar spectropolarimetric inversions.

The MC method described in this section has the potential to uncover possible ambiguous solutions, as the model can, in principle, converge to multimodal distributions. However, it is important to note that this approach is inherently limited by the assumption that the thermodynamic structure of the medium remains fixed. Fully uncovering all ambiguous solutions would require extending the MC procedure to explore the entire parameter space, including variations in the thermodynamic structure. While this would provide a more comprehensive analysis, it would also significantly increase the computational cost.

The level of noise in the observations impacts the quality of the inversion. In this study, we have not explored a range of different noise levels, as doing so would have introduced a large number of additional scenarios and potentially diluted the primary focus of the paper. However, a comprehensive analysis of inversion performance under varying noise conditions, along with the closely related issue of ambiguities in the inferred magnetic field, will be addressed in future work.

\section{Numerical experiment
\label{sec:experiment}}

To evaluate the performance of the solenoidal inversion, we constructed a forward model characterized by a simple thermal and density structure, combined with a relatively intricate magnetic field vector. Using this model, we synthesized the emergent spectra, introduce noise, and assess the inversion's ability to recover the underlying magnetic field configuration. We used the approach outlined in Sect.~\ref{sec:solen} to ensure the solenoidality condition and the method described in Sect.~\ref{sec:err} to estimate the uncertainty in the inferred magnetic field vector.

\subsection{The forward model
\label{sec:fwd}}

To establish the simplest possible thermal structure for benchmarking, we used a constant-property box representing a prominence-like configuration. In Fig.~\ref{fig:fov-big} we show the line-center intensity image in a FOV including the underlying solar disk. The computational domain spans $5 \times 5 \times 5 \, \mathrm{Mm}^3$, with uniform temperature, $T = 10^4\,\mathrm{K}$, and atomic number density, $N\,[\mathrm{cm}^{-3}]$ such that $\log N = 2.1$. The magnetic field is arbitrary and not intended to replicate any realistic prominence magnetic field topology. To evaluate the method's capability to infer the magnetic field vector and quantify the associated uncertainties, we chose the following vector potential components,\footnote{For notation simplicity, we disregard the normalization factor $\mu(A)$ in these expressions.}
\begin{align}
A_x&=T_2(y) T_2(z)\,,\\
A_y&=3 T_1(x) (T_1(z) - T_2(z))\,,\\
A_z&=T_2(y) - T_2(x)\,,
\end{align}
and the corresponding magnetic field vector components,
\begin{align}
B_x&=-3 x + 4 y\,,\\
B_y&=4 x + 4 (2 y^2-1) z\,,\\
B_z&=3 (z-1) - 4 y (2 z^2-1)\,.
\end{align}
The normalized Cartesian coordinates $x$, $y$, and $z$, ranging from the interval $[-1, 1]$, are mapped to the real domain coordinates during practical calculations. In the top panel of Fig.~\ref{fig:3dfield} we show the 3D structure of the model's magnetic field vector.

The model atom chosen in this study is purely academic, represented by a two-level, normal Zeeman triplet line. The critical Hanle field is set to $B_{\rm H}=284\,\mathrm{G}$, with an upper-level Land\'e factor of $g_u = 1$, an atomic mass of $2\,m_{\rm u}$, and a line-center wavelength of $5000\,\mathrm{\AA}$. The domain's line-center optical thickness is $\tau = 1.5$. For boundary illumination, we used the center-to-limb intensities at this wavelength from \citet{1976asqu.book.....A}, which account for the spherical geometry of the underlying Sun.

The synthetic observation consists of a $64 \times 64$ pixel grid, with added noise characterized with a standard deviation of
$\sigma = 2 \times 10^{-9} \, \mathrm{erg\,cm^{-2}\,s^{-1}\,Hz^{-1}\,sr^{-1}}$, which corresponds to
approximately 
$2 \times 10^{-4}$ fraction of the line-center intensity of the prominence. The FOV for this observation is shown in Fig.~\ref{fig:fov}.

\subsection{Inversion
\label{ssec:inv}}

The inversion process begins with an initial guess with randomized values for temperature and density and zero magnetic field. We modeled the spatial variation of these quantities with a set
of basis functions of order 4 along all Cartesian axes. Randomized angular quadratures from \citet{2020A&A...636A..24S} of order $L=15$ are used, following the method described in \citet{2022A&A...659A.137S}. The solution was obtained in approximately 16.5~kh of CPU time via a massively parallel run of POLARIS. The final inversion yields a value of $\chi^2 = 1.03$.

Although the required CPU time is significant, the information gained from the 3D inversion, which accounts for both Zeeman and Hanle effects, far exceeds what can be obtained using 1D methods. In the bottom panel of Fig.~\ref{fig:3dfield}, we show the inferred magnetic field vector.

In Fig.~\ref{fig:corr} we show correlation plots that compare the Cartesian components of the magnetic field in the inverse and true models at $10^5$ randomly chosen points in the model's domain. Despite the small optical thinness of the medium (the smaller the optical thickness, the less significant the expected radiative coupling between different regions), the Pearson correlation coefficients for all components exceed 0.9, indicating a significant accuracy of the inversion.

Following the strategy outlined in Sect.~\ref{sec:err}, we used an MC approach to estimate the uncertainties of the inferred magnetic field. We generated 100 magnetic field models, each with a randomized vector potential. The randomization is achieved by perturbing all the basis function coefficients, with a variance of $\sigma^2 = 0.3^2$ in the units of the normalization factor $\mu(A)$. All 100 inversions converge to a $\chi^2$ value below 1.5, and thus all of them are included in the subsequent analysis.

Next, we analyzed the uncertainties in the magnetic field inversion. Figure~\ref{fig:hist} shows histograms of the magnetic field components resulting from the ensemble of randomized models at a specific coordinate in the domain. The peaks of the distributions closely align with the true values of the magnetic field component, something that holds true for most regions of the domain.

In Fig.~\ref{fig:los} we show the variation of the magnetic field components along the LOS for four pixels in the FOV. The inverse model accurately reproduces the behavior of the true magnetic field vector. With few exceptions, the true values lie within the uncertainty intervals provided by the MC method.

Figure~\ref{fig:slices} shows the quantity 
\begin{equation}
U_{ij}=\sqrt{\frac{\sigma_i^2+\sigma_j^2}{\langle B_i^2\rangle+\langle B_j^2\rangle}} ,
\label{eq:uij}
\end{equation}
which is computed in the $ij$ plane, where $i$ and $j$ represent the $x$, $y$, or $z$ components of the magnetic field for three slices passing through the domain that intersect its central point. The relative uncertainty is typically large in regions where the magnetic field is weak, as expected, since these areas do not leave relevant fingerprints in the spectral line polarization. Visual comparison between the true (left column) and inverse models (central column) reveals that, in regions with large relative uncertainty, the error in the inverted field vector is noticeably larger than in regions with small relative uncertainty, proving the suitability of the quantity $U$ to describe the uncertainty.

\section{Discussion and conclusions
\label{sec:concl}}

We have presented a method for ensuring the solenoidality of the magnetic field vector inferred in 3D inversions, as well as for providing an estimation of the uncertainty of the inverted magnetic field. One important consideration when using the method is the degree of smoothness in the inferred magnetic fields. While this constraint may prevent the retrieval of small-scale turbulence, we think it is preferable to obtain a globally smooth solenoidal field rather than an arbitrary local field with uncertain solenoidality.

We have found that even a modest optical thickness of about 1.5 along the LOS is sufficient to provide high-quality magnetic field information at a given noise level. This finding is consistent with the results by \citet{2022A&A...659A.137S}, which emphasize the importance of optical thickness for the accuracy of magnetic field retrievals in self-consistent models.

In our numerical example, the magnetic field is on the order of the Hanle critical field, which ensures good sensitivity to the Hanle effect. However, an interesting research problem for the future will be exploring how the method behaves when the spectral line enters the Hanle saturation regime. This scenario may provide valuable insights into the potential ambiguities associated with the saturated Hanle effect.

Although we did not encounter any ambiguous solutions in our thermodynamically simplest possible model, the issue of ambiguities in the magnetic field inference remains a key topic for future research. This will include a theoretical investigation of the role of noise in the observation and the plasma's optical thickness in generating such ambiguities. Additionally, the uncertainties in the thermodynamic quantities, which were not addressed in this work, warrant further investigation to assess their impact on the inversion process.

Our immediate plans involve applying POLARIS to real observational data, focusing on structures in the upper solar atmosphere, including both prominences and the chromosphere. These regions often exhibit small to very large optical thicknesses, providing a rich domain for testing the applicability of our method in more complex and realistic settings.

\begin{acknowledgements}
J.\v{S}. acknowledges the financial support from project \mbox{RVO:67985815} of the Astronomical Institute of the Czech Academy of Sciences.
T.P.A.'s participation in the publication is part of the Project RYC2021-034006-I, funded by MICIN/AEI/10.13039/501100011033, and the European Union ``NextGenerationEU''/RTRP. T.P.A. acknowledges support from the Agencia Estatal de Investigación del Ministerio de Ciencia, Innovación y Universidades (MCIU/AEI) under grant ``Polarimetric Inference of Magnetic Fields'' and the European Regional Development Fund (ERDF) with reference PID2022-136563NB-I00/10.13039/501100011033.
We acknowledge RES resources provided by Barcelona Supercomputing Center in MareNostrum5 to the activity \mbox{AECT-2024-2-0019}.
\end{acknowledgements}

\bibliographystyle{aa}
\bibliography{ms.bib}

\end{document}